# Assessing inner boundary conditions for global coronal modeling of solar maxima

*Michaela Brchnelova*, michaela.brchnelova@kuleuven.be
*Center for mathematical Plasma Astrophysics, Department of Mathematics, KU Leuven, Belgium*

*Błażej Kuźma[1], Fan Zhang[1], Barbara Perri[2], Andrea Lani[1], Stefaan Poedts[1,3]*
[1]*Center for mathematical Plasma Astrophysics, Department of Mathematics, KU Leuven, Belgium*
[2]*Université Paris-Saclay, Saclay, France*
[3]*Institute of Physics, University of Maria Curie-Skłodowska, Poland*

**Abstract**

Computational Fluid Dynamics (CFD)-based global solar coronal simulations are slowly making their way into the space weather modeling toolchains to replace the semi-empirical methods such as the Wang-Sheeley-Arge (WSA) model. However, since they are based on CFD, if the assumptions in them are too strong, these codes might experience issues with convergence and unphysical solutions. Particularly the magnetograms corresponding to solar maxima can pose problems as they contain active regions with strong magnetic fields, resulting in large gradients. Combined with the approximate way in which the inner boundary is often treated, this can lead to non-physical features or even a complete divergence of the simulation in these cases. Here, we show some of the possible approaches to handle this inner boundary in our global coronal model COolfluid COrona uNstrUcTured (COCONUT) in a way that improves both convergence and accuracy. Since we know that prescribing the photospheric magnetic field for a region that represents the lower corona is not entirely physical, first, we look at the ways in which we can adjust the input magnetograms to remove the highest magnitudes and gradients. Secondly, since in the default setup we also assume a constant density, here we experiment with changing these values locally and globally to see the effect on the results. We conclude, through comparison with observations and convergence analysis, that modifying the density locally in active regions is the best way to improve the performance both in terms of convergence and physical accuracy from the tested approaches.

Keywords: magnetohydrodynamics, solar physics, solar corona, Computational Fluid Dynamics, numerical modeling

**Introduction**

Space weather forecasting through established numerical tools such as EUHFORIA (Pomoell and Poedts, 2018; Poedts et al., 2020) and ENLIL (Space Weather Prediction Center 2022) currently rely on a WSA-like coronal model. This PFSS-based model is computationally efficient, but often yields insufficient accuracy in predictions of the background solar wind, especially for fast wind streams. For that reason, COCONUT - the 3D global MHD coronal model was recently developed in order to improve this capability (Perri and Leitner et al., 2022). This code is based on an unstructured mesh and an implicit scheme, i.e. two features which allow it to operate faster than other state-of-art coronal modeling codes, as demonstrated in Perri and Leitner et al. (2022), where the detailed comparison with the explicit-in-time model Wind-Predict that is based on the PLUTO code (Mignone et al. 2007) is presented. This makes COCONUT suitable and desirable for operational purposes. However, despite being physically more accurate than WSA, it still relies on many approximations, some of which may significantly affect the accuracy of the solution and even the ability of the code to converge, especially in cases of high magnetic activity. The latter are especially important since the major space weather events happen around this phase of the solar cycle. For this reason, in this paper, we perform a numerical experiment in which we assess different modifications to the default inner boundary condition setup in order to improve the modeling of the corona at the maximum of the solar activity.

There are several approximations that we make on the inner boundary. First of all, we prescribe the magnetic field according to a photospheric/chromospheric magnetogram, despite the fact that the region represented by the inner boundary is located in the lower corona. This is due to the fact that we do not have capability yet to measure the global coronal field for the whole sun on a daily basis. It is likely that the real magnetic field in this region is different, with some of the high values and gradients potentially dissipated. Secondly, we also currently assume a constant density and pressure everywhere on the surface, even in active regions. That, again, is not physical as we know from local observations that the density and temperature in active regions should be higher, see e.g. the analysis of the data from Hinode and the work of Tripathi et al. (2008). Thus, the magnetic field and the density are the boundary conditions that we will modify in this study.

**Materials and methods**

In this work, we make use of the newly developed global coronal COCONUT model. The numerical framework, initial setup, and code verification are extensively discussed in Perri and Leitner et al. (2022), while parametric studies regarding magnetograms are presented in Perri et al. (2022). The detailed comparison with observations is shown in Kuźma et al. (2022). The code is, for now, polytropic and solves the set of ideal MHD equations with gravity:

$$\frac{d}{dt}(\rho) + \nabla \cdot (\rho \boldsymbol{v}) = 0$$

$$\frac{d}{dt}(\rho \boldsymbol{v}) + \nabla \cdot (\rho \boldsymbol{v}\boldsymbol{v} + \bar{I}[p + \frac{1}{2}|\mathbf{B}|^2] - \mathbf{B}\mathbf{B}) = \rho \mathbf{g}$$

$$\frac{d}{dt}(\mathbf{B}) + \nabla \cdot (\boldsymbol{v}\mathbf{B} - \mathbf{B}\boldsymbol{v} + \bar{I}\phi) = 0$$

$$\frac{d}{dt}(e) + \nabla \cdot ([e + p + \frac{1}{2}|\mathbf{B}|^2]\boldsymbol{v} - \mathbf{B}[\boldsymbol{v}\cdot\mathbf{B}]) = \rho \mathbf{g}\cdot\boldsymbol{v}$$

$$\frac{d}{dt}(\phi) + \nabla \cdot (V^2_{ref} \mathbf{B}) = 0$$

In the equations above, ρ is the density scalar, $\boldsymbol{v}$ the velocity vector, **B** the magnetic field vector, **g** the gravitational acceleration vector, *e* the internal energy scalar, p the pressure scalar, ɸ a divergence cleaning parameter, $V_{ref}$ the reference velocity constant for hyperbolic divergence cleaning method and Ī the identity matrix.

Inclusion of further, more sophisticated physical terms, representing the radiation, coronal heating and conductivity is the focus of the ongoing work. We utilize an icosphere-based grid of approximately 1.5 million elements (see Brchnelova et al. (2022b) for more information about grid effects) with the default boundary conditions derived by Brchnelova et al. (2022a). The default inner boundary setup prescribes:

- the magnetic field, the radial component of which is derived from the magnetogram (by default the HMI magnetogram is used, see Perri et al. (2022), with $l_{max}$ = 30. The $l_{max}$ of 30 parameter refers to the maximum spherical harmonic degree used for the reconstruction of the magnetogram field);
- the velocity, a default outflow of 1935.07 m/s aligned with the magnetic field;
- a constant density of 1.67e-13 kg/m$^3$ and a constant pressure of 0.039 Pa;
- the divergence cleaning parameter φ of 0.

In this study, we will manipulate these boundary conditions to improve the computational performance and physical accuracy of the results. In the work of Kużma et al. 2022, it was presented that for the cases of solar maxima where the magnetic field values and gradients exceeded certain limits (which were dependent on the resolution of the magnetic map, the computational grid and the limiter applied), convergence issues might appear, giving rise to negative local temperatures and pressures. Since this behavior was only observed for the cases of solar maxima, a solar maximum magnetogram from November 2012 (Carrington Rotation 2130) was selected for the study. The date of the solar eclipse in that month was chosen since for this date (November 13), we also have high resolution observations of the solar corona, in this case provided by C. Emmanoulidis and M. Druckmüller[1]. In addition, during this time period, the Sun was very active and there were many visible different streamers at various latitudinal and longitudinal angles, which makes it a very challenging case that could not be converged with the default setup. Some of these features were also located in the polar regions which can create further convergence issues and disagreements with observations, since as seen in the previous study (Perri et al. 2022), the resulting flow field is fairly sensitive to the way in which the poles are resolved. Corresponding observations are shown on the left-hand side of Figure 1. On the right-hand side, the respective HMI magnetogram ( from NASA SDO) is presented.

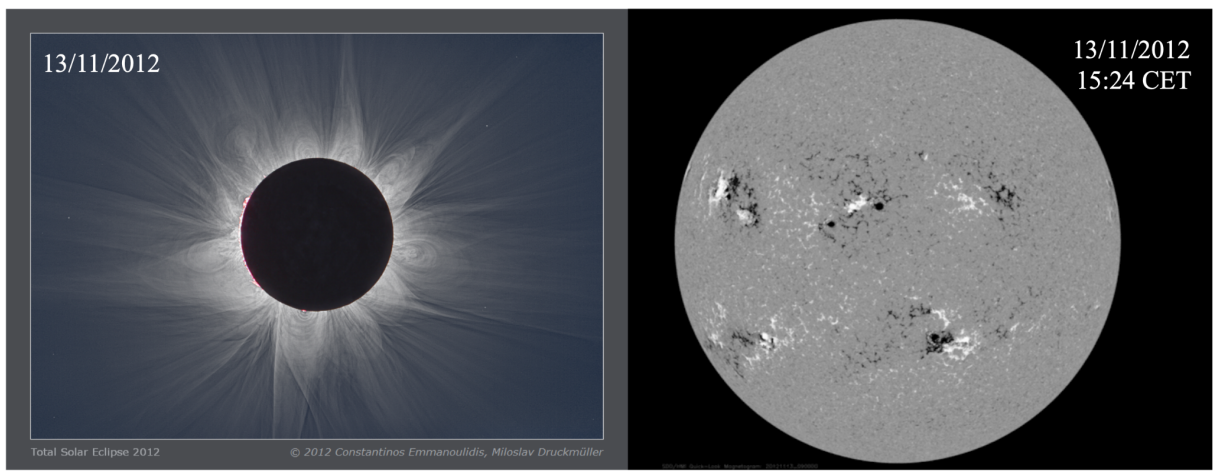

*Figure 1. The observation of the solar eclipse during a period of solar maximum (left), showing a variety of streamers in different directions, provided by C. Emmanoulidis and M. Druckmüller and the respective HMI magnetogram from NASA SDO (right).*

It was indeed observed that for this specific case, while the simulation was converging, non-physical negative temperatures and very high-speed streams developed in the domain above the active regions. This particular behavior is outlined in more details in Kuźma et al. (2022).

---

[1] http://www.zam.fme.vutbr.cz/~druck/eclipse/Ecl2012a/0-info.htm

In this work, to eliminate the above-described behavior, two different types of techniques were applied. The first type focused on altering the input magnetogram in order to reduce the magnetic field maxima and gradients by means of:

1. reducing the resolution from $l_{max}$ = 30 to $l_{max}$ = 15 and,
2. clipping out the maximum values of the magnetic field before processing with spherical harmonics.

What value would be used as the threshold in the second case was determined to be the value with which we no longer observed any negative temperature within the computational domain.

The second approach was to modify the prescribed density:

1. globally and uniformly, which is known to have beneficial effects on convergence,
2. and locally, in the regions where magnetic activity was strong.

In the first case, the smallest density enhancement that allowed the simulation to converge was multiplying it by a factor of 8. This threshold was selected due to the fact that with this value, all other tested cases with high magnetic activity could converge without the high-speed stream regions developing. In the latter case, the active region density was constrained by prescribing a maximum Alfvén speed. In order to allow faster convergence, a smooth hyperbolic tangent function was used to gradually increase density in the regions where the local Alfvén speed exceeded 2 million m/s. Here, the value was chosen such that it would be the least constraining value at which negative temperatures and pressures were not observed. To determine where this threshold lies, values in the range between 500 000 m/s and 7 million m/s were tested.

It would be of course possible to modify these parameters in combinations, for example, increasing the density only slightly while reducing the strength or the resolution of the surface magnetic field, but to a lesser extent than what was shown here. It is, however, easier to trace all the resulting uncertainties and possible errors in the simulation outcome when only one of such adjustments is applied. Thus for now, these methods are applied independently. In the future, when optimizing the performance of COCONUT or trying to converge especially difficult cases, it is possible that these methods will be used in tandem.

**Results**

In order to determine which of the techniques performs the best, out of the four approaches which were presented in the previous section, we looked at two kind of parameters:

I. the alignment of the streamers with the observed ones for physical accuracy and,
II. the number of iterations per second and convergence to a given threshold residual for numerical performance,

as both of them are important for space weather forecasting.

From the observational data, we determined the directions of the most prominent streamers and indicated them in blue. This is shown in Figure 2), where the rotation is due to the tilt of the Earth axis to bring the observation into the same frame as the simulations. .

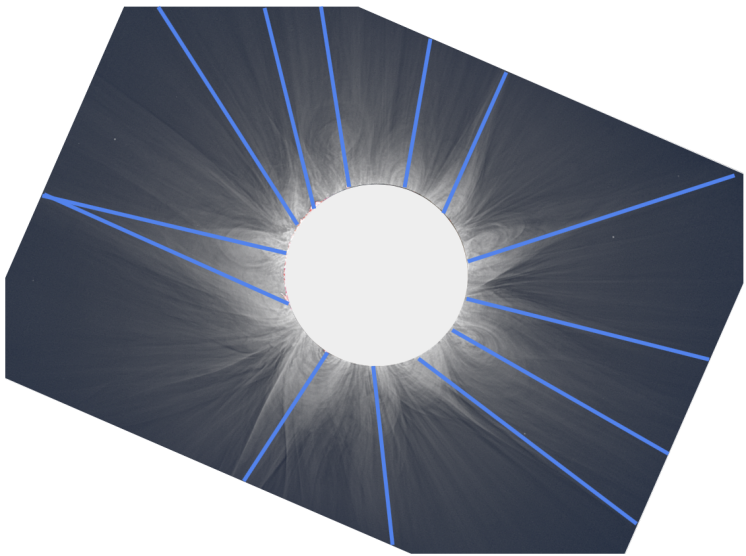

*Figure 2. The determination of the directions of the most prominent streamers from the observation shown as blue lines, with the photo rotated to correct for the Earth's tilt.*

From the simulation results, the same was done based on the magnetic field lines with the streamer directions shown in red. The corresponding results are shown in Figure 3.

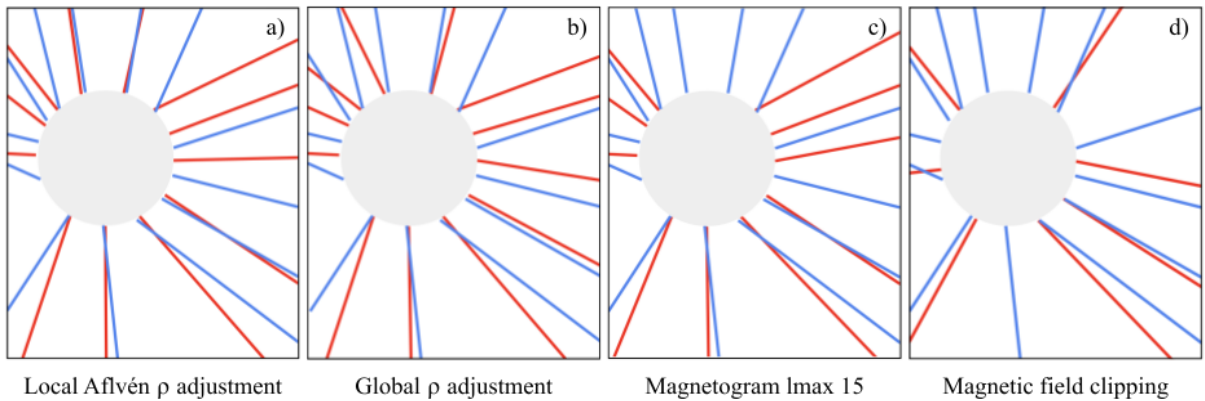

*Figure 3. The comparison between the streamer directions from the simulation (red) and from the observation (blue) for the four different simulated boundary conditions.*

The lowering of the resolution of the magnetogram and cutting off of the maximum absolute values prevented the model from being able to resolve the streamers near the poles (see the c) and d) panels of the Figure 3). There were not such significant differences in how well the streamers were resolved between the two density adjustment techniques, but the one where the density is adjusted locally (the a) panel of the Figure 3) instead of globally (the b) panel of the Figure 3) generally improves the alignment with the observed streamer direction. This is actually expected. As stated in the introduction, it is known that the density and temperature in active regions should be higher and thus this approach should be physically superior.

From the operational standpoint, all of the simulations were executed at a constant CFL of 2 for the same amount of CPU-hours on the system (the Genius cluster of the Vlaams Supercomputer Centrum). The convergence history for the four techniques is shown in Figure 4)*.* It is clear that all of the cases achieve roughly the same residual at 2500 iterations. In addition, roughly the same number of iterations was reached in the given CPU-hours, meaning that the time per iteration for each of them was similar. Operational performance thus does not play a role in the selection of the most suitable technique, which thus remains the local density enhancement due to its superior accuracy demonstrated above.

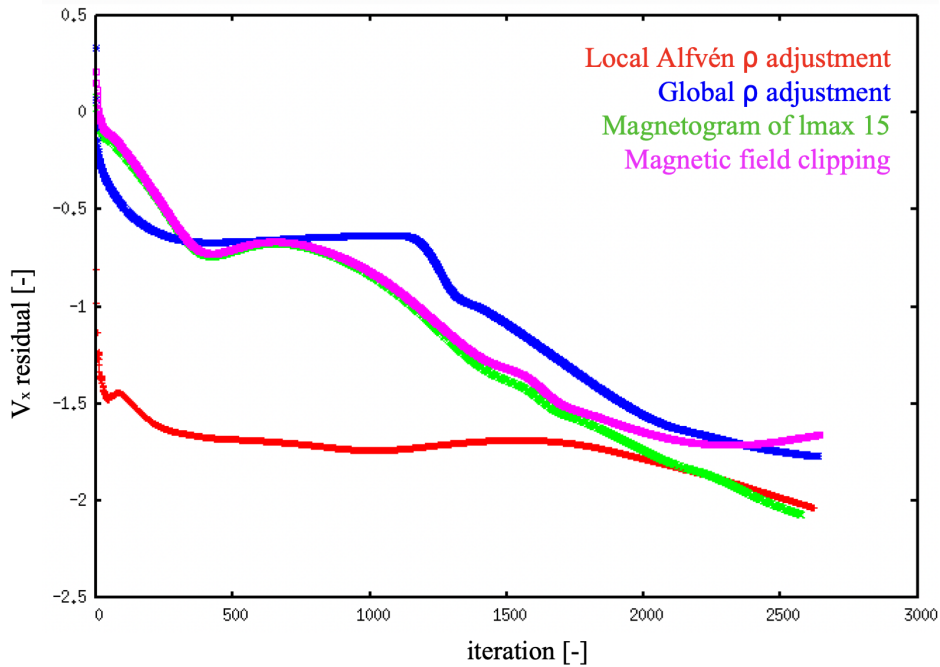

*Figure 4. The convergence history of the four simulated cases for the same CPU-hours of runtime, showing a similar residual after 2500 iterations and similar time per iteration.*

Here, only one case was presented. Based on what we have observed with other cases however, we expect the behavior to be at least somewhat similar for the following reasons. The fact that using a lower resolution map and/or clipping the maximum magnetic field values dissipates the strongest features, helps convergence but limits the result accuracy was already reported for a different maximum case in Kużma et al. 2022. In addition, we have seen that the zones with locally high Alfvén numbers are where the negative temperatures and pressures are located and where the simulation is most likely to diverge. There were several such zones above various active regions in the original case, and by reducing the maximum Alfvén speed ( by increasing the density either locally or globally), all of these were eliminated, regardless of the position or nature of the active region. Thus it is reasonable to expect such a behavior also with other active regions in magnetic maps from different dates.

**Discussion**

From the results presented above we infer that the local density enhancement according to a cut-off Alfvén speed is, for the case presented, the best technique to allow for efficient and accurate maxima modeling within the framework of the COCONUT code. This approach, however, must be thoroughly tested before this claim can be generalized on the global coronal models. We propose this technique to be applied to more maxima cases with aim to i) determine whether it indeed remains the superior technique in terms of physical accuracy; ii) prove that it can remove convergence problems for maxima universally, and iii) evaluate whether the selected cut-off Alfvén speed of 2 million m/s can be also used universally with these settings. If not, and if different cut-off Alfvén speeds are required for achieving a good performance for different maxima cases, this approach will not be suitable for space weather forecasting from the operational standpoint as it will not be possible to automatize. Instead, the density should be increased globally for numerical stabilization, or another technique should be developed. Thus, testing the above described techniques on more magnetograms is the essential step to further take before the model COCONUT can be exploited as a space weather prediction tool for cases of solar maxima.

**Acknowledgements**

This work has been granted by the AFOSR basic research initiative project FA9550-18-1-0093. This project has also received funding from the European Union's Horizon 2020 research and innovation program under grant agreement No.~870405 (EUHFORIA 2.0) and the ESA project "Heliospheric modelling techniques" (Contract No. 4000133080/20/NL/CRS). These results were also obtained in the framework of the projects C14/19/089 (C1 project Internal Funds KU Leuven), G.0D07.19N (FWO-Vlaanderen),

SIDC Data Exploitation (ESA Prodex-12), and Belspo project B2/191/P1/SWiM. The resources and services used in this work were provided by the VSC (Flemish Supercomputer Centre), funded by the Research Foundation - Flanders (FWO) and the Flemish Government.